# DEVELOPING A FUNCTIONAL PROTOTYPE MASTER PATIENT INDEX (MPI) FOR INTEROPERABILITY OF E-HEALTH SYSTEMS IN SRI LANKA

Dr Jayathissa W.G.P.T (Post Graduate Institute of Medicine University of Colombo)
Prof Vajira H W Dissanayake (Post Graduate Institute of Medicine University of Colombo)
Dr Roshan Hewapathirana ((Post Graduate Institute of Medicine University of Colombo)

## ABSTRACT

*Introduction: A Master Patient Index(MPI) is a centralized index of all patients in a health care system. This index is composed of a unique identifier for each patient link to his/her demographic data and clinical encounters. A MPI is essential to ensure data interoperability in different health care institution. The Health ministry of Sri Lanka planning to develop MPI for the country. This project focused on developing the prototype MPI for Sri Lanka with the view to implementing and scaling up at national level.*

*Methods: This project consisted of 3 phases. Phase 1: requirement analysis using focus group discussions (FGD) with information system users. Phase 2: identification of the suitable Application Programming interface (API) model. Phase 3: development of the prototype MPI.*

*Results: FGD were conducted in 6 hospitals. There were 78 interviewers (Male -36, and female - 42). They highlighted the key requirements for the MPI. Which were the unique identification method and different searching criteria and merging records to avoid duplication. Using this information, the requirements specification for MPI was developed. A combination of mono lithic and micro services architecture was selected to develop the MPI. The API using the Personal Health Number (PHN) as the unique patient identifier and HL7 standard was developed and implemented.*

*Conclusions: Development and implementation of a MPI has facilitated the long due need for interoperability among health information systems in Sri Lankan.*

## KEYWORDS

*MPI, Interoperability, Unique Identifier, PHN, API*

# 1. INTRODUCTION

## 1.1 Interoperability

"The ability of two or more systems or components to interchange information and use predictably the information that has been exchanged"[1]. Interoperability means the ability to communicate and exchange data accurately, effectively, securely and consistently with different information systems, Software application and networks in various settings and exchange data such that clinical or operational purpose and meaning of meaning of the data are preserved and unaltered[1]. Some do work with interoperability within a hospital or clinic or from one department, such as the emergency department to another. Interoperability archived through HL7 and XML basically. Health Level 7 (HL7) is an organization that develops and defines standards to facilitate communication between systems and applications are linked to the health area[2].XML is a software and hardware independent tool for transporting data and storing data. For the semantic interoperability XML is extensively used. For the interoperable system this above two methods, HL7 and XML activity take part in exchanging data with other systems.[3]

## 1.2 Master Patient Index

A Master Patient Index (MPI) is an electronic medical database that holds information on every patient registered at a healthcare organization. It also includes data on physicians, other medical staff, and facility employees. Master patient index is a database that is used all over the world in healthcare organization maintains consistent, accurate and current demographic and essential medical data of the patients seen and managed within its various departments[4]. The patient is assigned a unique identifier that is used to refer to this patient across the enterprise. MPI is a form of customer data integration which is implemented by Healthcare organizations for identify, duplicate, match, and cleanse and merge patient records. After creating a MPI and implementing will used to obtain a complete and single view of a patients. "The MPI create a unique identifier (UI) for each and every patient and maintain it with the mapping to the identifiers used in each record's respective system"[5]

Health Information system is marching diligently toward a more connected system of care through the use of EHRs and electronic exchange of patient information between hospitals[6]. The Patient Identification and matching are focused on to help ensure the accuracy of every patient's data and the availability of their information wherever and whenever care is needed[7]. From the respective systems, MPI provides an Application Programming Interface (API) for querying and searching the MPI to find patients and the ways or the pointers to their identifiers and records. Also, store some subset of the attributes and data for the patient so that it may be queried as an authoritative source of the "single most accurate record" or "source of truth" for the patient[4]. Registration or other practice management applications may interact with the index when admitting new patients to have the single best record from the start or may have the records indexed at a later time.

Application programming interfaces make it easier for developers to use certain technologies in building applications for the interoperability of health information systems. By abstracting the underlying implementation and only exposing objects or actions the developer needs, an API reduces the cognitive load on a programmer. Reduction of cognitive load increases the functionality of the programmer's program to overcome defeats like bugs. While a graphical interface for clients might provide a useful with a button that performs all the steps for fetching and highlighting new contents, an API for file input/output might give the developer a function that copies a file from one location to another location.

## 1.3 Sri Lankan Health care setting

Government health care system has several service deliveries, such as administration, training, curative care, public health, other resource management programs[9].

In Sri Lanka, curative care is the Hospitals mainly categorized into several levels according to resources allocation and resource availability to the institutions. Some of them belong to the line ministry. Line ministry is the central core or the backbone of the health ministry. The Hospital that attached to the line ministry are the teaching hospitals, provincial general and district general hospitals, especial care unit like cancer institution and the rehabilitation units. Other hospitals are attached to the provincial council under the provincial health ministry. Inside the curative sector, there is a referral order from the lower level of curative care (primary care institutions) to the upper level (tertiary care institutions). It is important to maintain the sustainability of this system to provide better care for the bath paediatric population and the adult population without time delays and lags. The public sector provides 95%of inpatient care while private sector responsible for remaining 5%. However, this referral system does not prevent from directly going to private practices by consultants and general practitioners as well as admission to private hospitals[8].

## 1.4 Interoperability of Health Information Systems in Sri Lankan contest.

Sri Lanka is a developing country, which is establishing towards to the using of the technology of the new era. Comparative to other developing countries the science and technology is passed to Sri Lanka simultaneous with the western countries. With the globalization and free trading environment in Sri Lanka open the door to the western world to trade their science and technical products through the country. The low-cost product supports a lot to grow the enthusiasm on the people to fall on the track with the westerners where they used in day to day life. Electronic Medical Records systems (EMR) work as a standalone application in Sri Lanka. Government hospitals are established with HIMS/HHIMS (health Information management systems/Hospital health Information management systems) which is developed by a Sri Lankan software vendor ICTA (Information and Communication Technology Agency). At this moment, functioning of HHIMS/HIMS in the hospital without any interaction with the other existing electronic health record system. The interoperability among the systems is not established. To overcome this issue, the government should plan to develop some application that could overcome the interoperability issues.

Developing a MPI is the 1st step towards the interoperability of among standalone systems. By correctly matching patient records from disparate systems and different organizations, a complete view of patient records may help for the continuation of patient care. With this complete view, numerous benefits may be realized including a better patient care and continuation of the patient care pathway.Improved customer service can have offered with reduced waiting time at hospitals clinics and Outpatient department (OPD). In an emergency or other critical care situations, medical staff more confident that they know medical conditions or other health related information that would be critical to providing proper care. Past medical history, past surgical history and clinical care related information, can be obtained from across organizations. Blood group and essential minimal clinical data will provide unconscious patient to be live after 5 hrs.
other than death without intervention.

Need for a MPI in Sri Lankan context is that to bridge the gap between interoperability issues among the HHIMS/HIMS in Sri Lanka. MPI will establish the continuation of care pathway by multiple health care providers. Enable exchange of minimum essential data set among different EMR systems for the Sri Lanka. In continuation of care pathway or other critical care situations, medical staff can be more confident that they know medical conditions or other information that would be critical to providing proper care. Retrieval of past medical history, allergy history. Family history from multiple EMRs to the EMR at current care provider.

## 1.5 Objectives

### 1.5.1. General Objective

Developing a Functional Prototype Master Patient Index (MPI) for Interoperability of e-health Systems in Sri Lanka.

### 1.5.2. Specific objectives.

1. To Identifying and suggest solution for EMR interoperability through a prototype MPI. This including developing A high-level software architecture of MPI and formulate Functional Requirements for a MPI.

2. To propose suitable API model for the proposed MPI

3. To suggest essential message template for MPI

## 2. METHODS

## 2.1 Action Design Research (ADR)

For information system development has two basic missions to complete to achieve the successful product and implementation.1st mission is contribution to the theoretical aspect of the research and 2nd aspect is how to show the problems of the system using the current practices. For system development and implementation, the action research(AR) and the design research(DR) methods being used. In the design research, the new design is developed and the action research solves an immediate organizational problem. To make a product as successful one researcher has to put the theory into practice in a practical way. The design research and the action research work in a similar way. Designing the master patient index is a DR and the implementation in the health care setting is an AR. ADR method "IT artifact is built and evaluate in an organizational setting using a general prescriptive design" (10). The research project is a qualitative research. I used action design research(ADR) as the method for development of the MPI.

ADR method "IT artifact is built and evaluate in an organizational setting using a general prescriptive design" (10). This deals with 3 main areas of the IT artifact development. 1st problem formulation process using the design research (DR) method .2nd Evaluation using DR method. 3rd and final are the artifact building. As according to the ADR method research project is carried out in 3 phases. In the phase 1 problem formulation done. Phase 2 building, intervention and evaluation process for the MPI Architecture. At the final phase developing the system after formalization of what learn from the previous phases. In the phase 1 data collection using focus groups, discussion and analysis of its result will be done in order to identify data elements and user requirements of MPI prototype. Phase 2 involves in identify the suitable API model for the proposed MPI. Phase 3 is developing a prototype MPI using the identified user requirements.

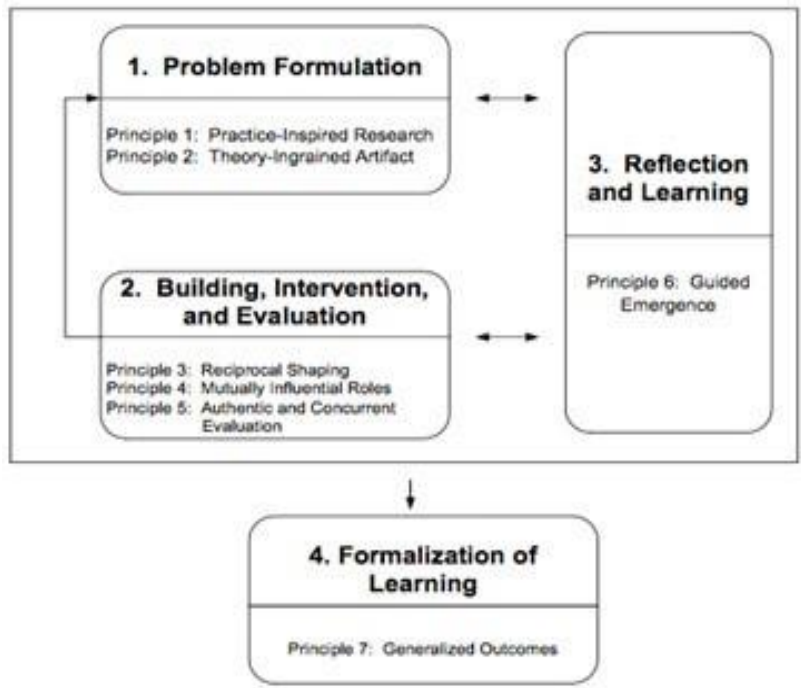

Figure 2. 1 ADR Method - Stages and Principles(10)

## 2.2 Phase 1

In the phase 1 did the problem formulation. During this phase practical, inspired research principal used. In this phase viewing the field problem and knowledge gathering and creating opportunities is the main target. Opportunity is the intersection of technical and organizational domains. Action design researcher will get the knowledge to apply to set of problems to solve the problem (10).

This was carried out using Qualitative research design. Formative evaluation be applied to the qualitative research design. The research did on the staff members who are involved in the use of the HHIMS/HIMS health information system in the relevant respective hospital of two districts (Colombo and Kalutara). The sampling method is Convenient sampling method is chosen mainly because of the limited resources and time. Study instruments Semi-Structured interviews for HHIMS/HIMS users. Focus group discussions and interviews with, HHIMS/HIMS users and Software Development group team. Interviews and Focus group discussion (FGD) conducted by the principal investigator with the help of supervisors. Data recording was done correctly. Randomly selection of participants/groups for FGD was done. FGD was done following an unstructured interview at the beginning, and then semi-structured questions was created to ask later by avoiding the unimportant issues arrived at the earlier conversation.

After taking permission from the relevant authorities the principal investigator introduced himself to the staff at the above relevant respective hospital of Colombo and Kalutara district and explain the research and the methodology to the eligible participants of the research. Informed consent was obtained from the eligible participants after explaining the full investigation procedure including benefits and risks. An adequate time allowed for them to clarify any query or further questions. Informing them that they can withdraw from the study at any time without giving any reason and no further correspondence made ensures voluntary participation.

Data from interviews analyzed using qualitative methods to find relevant concepts related to the development of MPI. Collected ideas recorded using computer software. These recoded data used to identify themes, Concepts, requirements of the users. Research did not involve collecting personal identification information of participants during data collection. Therefore, privacy and confidentiality of the participants are preserved. Ethical approval has taken by ethics review committee PGIM, University of Colombo. Before collecting data from health institutions necessary permission gained from the head of the institutes. Privacy, security, and confidentiality of data assured. The purpose of the research and publication would not involve using any personally identifiable data of anybody or there is no possible access the system or its data by third parties because all data gathering would be done by researcher's personal computer.

## 2.3 Phase 2

This phase included designing cycle of the IT artifact. For the development MPI design its architecture was built in this phase of the development of the MPI. Using the both domain of expertise gets together and worked out for a plan for an architectural design of the MPI. Three basic principles used in the development of the architectural design. The 1st did building of the architectural design. After the building of the design intervention made whether that is suitable for used. Evaluation was carried out at the end. The literature review of existing systems used as the tool for the evaluation.

Early design process created the Alpha-version of the MPI Architecture. After a thorough literature review and the discussion with the development team, the beta version of the designed.

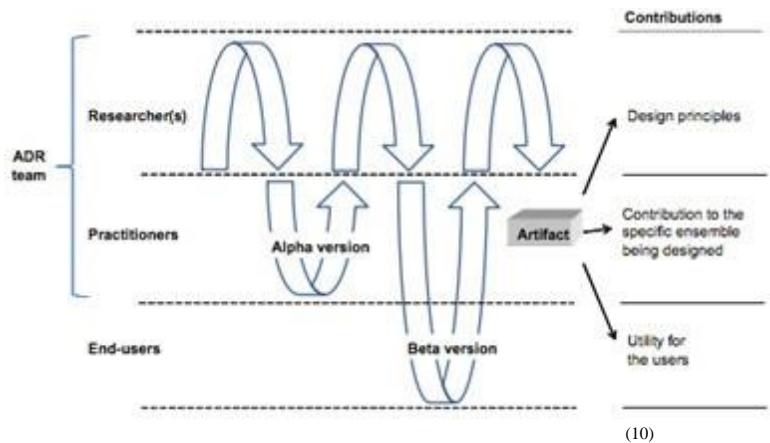

Figure 2. 2Development of The Alpha Version

During the period of evaluations through the literature reviews API model for the proposed MPI was identified. Researcher identified the core interaction and message types between electronic health record and master patient index. Devised a mechanism to incorporate unique identification number to MPI

## 2.4 Phase 3

After identifying the Architectural design and the set of APIs and the core interactions and message type architecture of the MPI the phase 2, Software Development life cycle started following Scrum software development practice. Scrum is an agile software development approach which delivered software programme deliverables at end a software development phase. This procedure is well known among the young software developers who used to develop web based programmes.

The software development team atprocessed of the software development as agile software development using scrum. Reason for using the scrum everything is visible to the researcher while researcher able to track down what development process complete up to now and what process should elaborate more time for the development. Development will be piece by piece so at every 2 weeks researcher will have product on his hand.

Scrum procedure for the development of the MPI in Phase 3.

1. Create a backlog: All the feature the researcher like to have contained on this document. Basically, the researcher's wishes and the dream of the project. This was created by the researcher using the phase 1 requirement document.

2. Estimate time then researcher prioritized the most important task as first and the least at the end. At the end of this process release backlog to the development team as a prioritized list.

3. Sprints as 2weeks period for each delivery of the task. At start of every sprit there was a meeting and discussed what task will be done at the end of the sprint. At the end of the sprint, deliverable given to the researcher and the plain for the next sprint.

Identified Scum master for the development procedure by the development team. After the development of the full requirement document this scrum procedure started. MPI is developed using the identified user requirements. Testing message architecture using Prototype MPI and HHIMS.

Through this development period all the team worked through tied time schedule to give the complete product to the researcher. Development team consisted of separate software development, software testing and software security teams. Procedure was 1st development then testing at the end security measures taken at each software deliverables.

## 2.5 Reflection and Learning

In the 21st century the information system diversity is immense and great. The use of these  greater diverse technologies in the filled of healthcare and medicine becoming more commonly. More and more usage of technologies in the different specialties in the healthcare domain[11]. These technologies are getting more advance and scalable. Nowadays there is a great diversity of information systems within all healthcare Providers. Each one with different specifications and capabilities as well as communication methods, thus hardly allowing scalability. This heterogenic Set of characteristics is an impediment to achieve interoperability between systems which, indirectly, affects the patients' well-being[12]. It is common that, upon watching Each database of each one of these information systems, researcher noticed different entries referencing the same person entries with insufficient or wrong data, due to errors or miscomprehension upon patient data insertion and outdated data[13]. These problems bring duplicity, incoherence, lack of actualization and dispersion on patient data. It is with the purpose of minimizing these problems that the Master Patient Index Concept is needed. A Master Patient Index proposes a centralized repository, to index all patient entries within a pre-determined set of information systems. This repository Is composed by a set of demographic data, sufficient to unmistakably identify a person, and a list of identifiers that unequivocally identify the several entries that a patient Possesses within each information system. This solution allows synchronism between All intervenient minimizing incoherence, lack of actualization, inexistent data, and Diminishing entry duplications[14]. The Master Patient Index is an asset not only to the patients but also to the medical Staff and healthcare providers.

In the current health facilities, there is a great diversity of systems of information. Each one with different specificities and capacities, Proprietary communications, and hardly allow scalability. This set of interoperability of all these systems, in the wealth of the patient. It is common practice that, when looking at all the databases of each Information systems, we come across different registers the same person; Records with insufficient data, records with incorrect data, due to errors or misunderstandings when inserting patient data and records with outdated data. These problems cause duplicity, inconsistency, and dispersion in patient data. It is with the intention of minimizing these problems that the concept of a Master Patient Index is required[15]. A Master Patient Index proposes a centralized repository, which Indexes all patient records of a given set of information. A repository which is constituted by a set of demographic data, Sufficient to unambiguously identify a person, and a list of identifiers, which identify the various records that the patient has in the repositories of each system of information. This solution allows synchronization between all the actors, mini-Incoherence, outdatedness, lack of data, and a Duplication of records[16].

For information system development has two basic missions to complete to achieve for successful product and implementation.1st mission is contribution to the theoretical aspect of the research and 2nd aspect is how to show the problems of the system using the current practices[17]. During the development of the prototype MPI researcher gather knowledge about the requirement for the MPI. During the designing phase researcher learn the basic design methods available for development as well as the what was being used in the past for the MPI development and implementation through a though literature review.

In the phase 2 of the development of the prototype MPI researcher studied about the monolithic system design and the micro service system design. Decision taken to use both micro service and monolithic service system design for the MPI architecture after the discussion with the technical experts of prototype development. Unique patient identifier, the PHN is introduced to MPI as Primary specific key for all integrations for other health domain systems. Developed the suitable API sets which enable the gateway for interoperability of health information systems.

Developing a MPI establishes interoperability of among HHIMS/HIMS systems in Sri Lanka. PHN as a unique identification number which integrated the laboratory information system(LIS), Picture archival and communication system(PACS), Citizen mobile application(CMP) to the public access. Clinical repository or the minimal clinical data set(MCDS) will be integrated to the PHN in the version 2 of the MPI. API sets enable interoperability of health information system.

## 3. RESULTS

### 3.1 Requirement Analysis

During the phase 1 researcher identified what are the system requirement for the development of MPI. The functional and non-function requirement identified then the document developed for the developer use.

FGD were conducted in 6 hospitals. There were 78 interviewers (Male -36, and female - 42). They highlighted the key requirements for the MPI. Which were the unique identification method and different searching criteria. All the interviewers were agreed about the unique identification number. During the discussions, the new concept of PHN introduced them and all were 100% agreed and excepted that the PHN is good method as a unique identifier for the MPI.

56 HHIMS users as percentage 71% of the interviewers agreed using of biometrics for the as searching criteria MPI. Some proposed fingerprint as biometric which is well used in the government offices and that would be a good form of method to identify unconscious patient. Some were suggesting facial recognition as searching criteria. 8 persons as percentage 10% were completely disagreed the biometrics. By their own words "*biometrics will violate patient privacy not need at the moment", "you should arrange some backend communication from the software for the future use of biometrics*". Some were not concern about biometrics at all.

All agree with good security and backup needed for the MPI. Breaching security measure affect the patient health care continuity. So, the good security protocols were concern irrespective of the cost. During the computer server failures, few suggest keeping all the data in the MPI will be a best option till full recovery. The health information system administrators were concerned about the introduction of cloud service to the MPI, so the server failure will support.

Duplication consider as the main topic that affect the clinical judgment. One doctor who handled the system for long duration brainstorm me with following facts which I have not think at all, he said that there is an option to save the previous drug prescription in the system and can reproduced it in a single click if wrong bar code or duplication can affect prescription and also allergies are visualized in red colour pop up message so the doctor no need to asked for allergic history again and can prescribe allergic drug to patient.

To avoid duplication suggested to merge the records using domain experts. Guardian assign for the children as well as for the mentally unsound patients and unconscious patient were considered by some interviewers. One interviewer suggested giving PHN to every live birth will reduce the back lock. Some suggested that verification of demographic data through population registry will be used full in future. Use of passport number for the PHN generation was consider for foreign patients. Elderly number is considered as a suggestion for searching criteria as well as for PHN generation.

11 HHIMS users as percentage 14% users agreed with the anonymity for the patients. Some says that "*in free health, seeking of anonymity do not happen*", one senior doctor said "*In my 21 year of my service I have worked in different place and I have never seen a patient who seek anonymity but they seek for privacy and confidentiality*", one postgraduate diploma holder who had worked at STD clinic for several years said "*In central STD clinic patients were given a card with a number that does not have any demographic data even not mentioned as STD clinic, all the investigation and drug are issued for that number*". But he also motioned that loss of the card is very common and duplication is common. He mentioned another scenario commonly happen, who were positive for VDRL and waiting to travel to middle east will come and give the history to the doctor and another person will carry the card and go to the lab and give the blood. The patient will get a negative result. He suggested if the is some form of identification method to prevent this happening. Some suggested "*only provide if somebody asked for anonymity*". In hospitals patients are coming with someone else bar code and take treatment. If any kind of a personal identification method were established in the system missed use of bar code never happen. Also, one person mentioned never violate Hippocratic Orth. Administrator said, "*by the constitution given the right to any citizen in the country to seek free health, please do not violate that right by introducing any unethical things*". One other administrator said government should provide sufficient fund to laminate the PHN card given to the patient, because the barcode print that now using is not durable.

## 3.2 MPI Architecture

After the completion of the phase 1, researcher learned the system requirement and started the designing phase of the architecture of MPI. During the discussion with the developers the 1st design architecture formulated. Monolithic type architecture is introduced as the 1st version of the MPI Architecture. If MPI developed on this architecture will be like a web application and problem was that the authentication of MPI and clinical repository on the same deployment version. When the outside requests are exceeding the limit, application will crash.

1st architecture is based on the monolithic service architecture and it will give the vendor a fast performance, but the authentication and other request are in the same application, when the 5000odd request come to the application from that of 5000-odd request the more than 90% request will be for the authentication. If developer available to remove the authentication part to a separate module to work, then we can get rid of above mention technical difficulties[18].

During the completion of the version 1 of the MPI came a conclusion that when considering above all the problems to introduce the Oath-Server to the system. Oath-Server work as a following in the figure.

By introduction of the OAuth server protocol increases the productivity of the MPI solution architecture. Oath server check the credibility of the authorization module feed back to the other services available processed[19]. By using an Oath-server reduce the separate logging twice to the systems when the in HHIMS logging to MPI to search for the clinical repository later. So, time consume for the separate authentication will increases the waiting time for the patient.

After the discussion with software developing team version 2 of the MPI architecture developed. This development is continuation of the same previous architecture. At the end of the discussion of 1st Architecture, introduced the Oath–server separately to the system design. During further discussion, the beta version introduced with more secured design. During the all discussions, the security was the utmost important thing is that the researcher learned. By adding the Oath – server to the architecture upgraded the system functionality by securing the data that restricting the unauthorized trans access. This architecture basically developed using the both monolithic and micro service Architectures

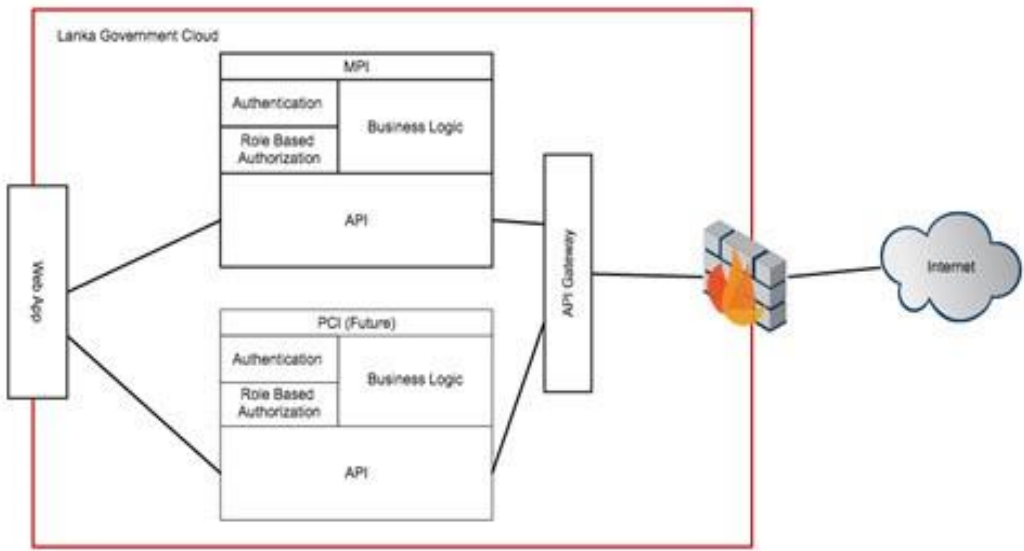

Figure 3. 1 Alpha Version of The MPI Architecture

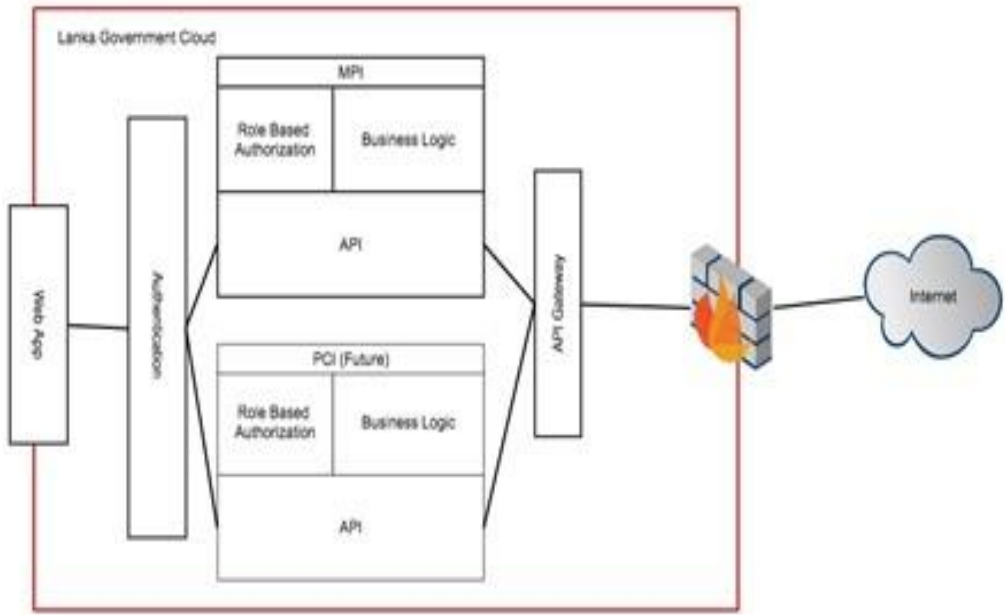

Figure 3. 2 Beta Version of The MPI Architecture

At the last version of the expanded MPI architecture is formulate for the purpose of the future public use with developing the mobile application which can be connected via API to the personal health data after dual authentication.

Researcher learned details about the procedure of the scrum and what is important and why is suitable for this developing this MPI. Use case diagrams formed for the better understanding of the MPI development.

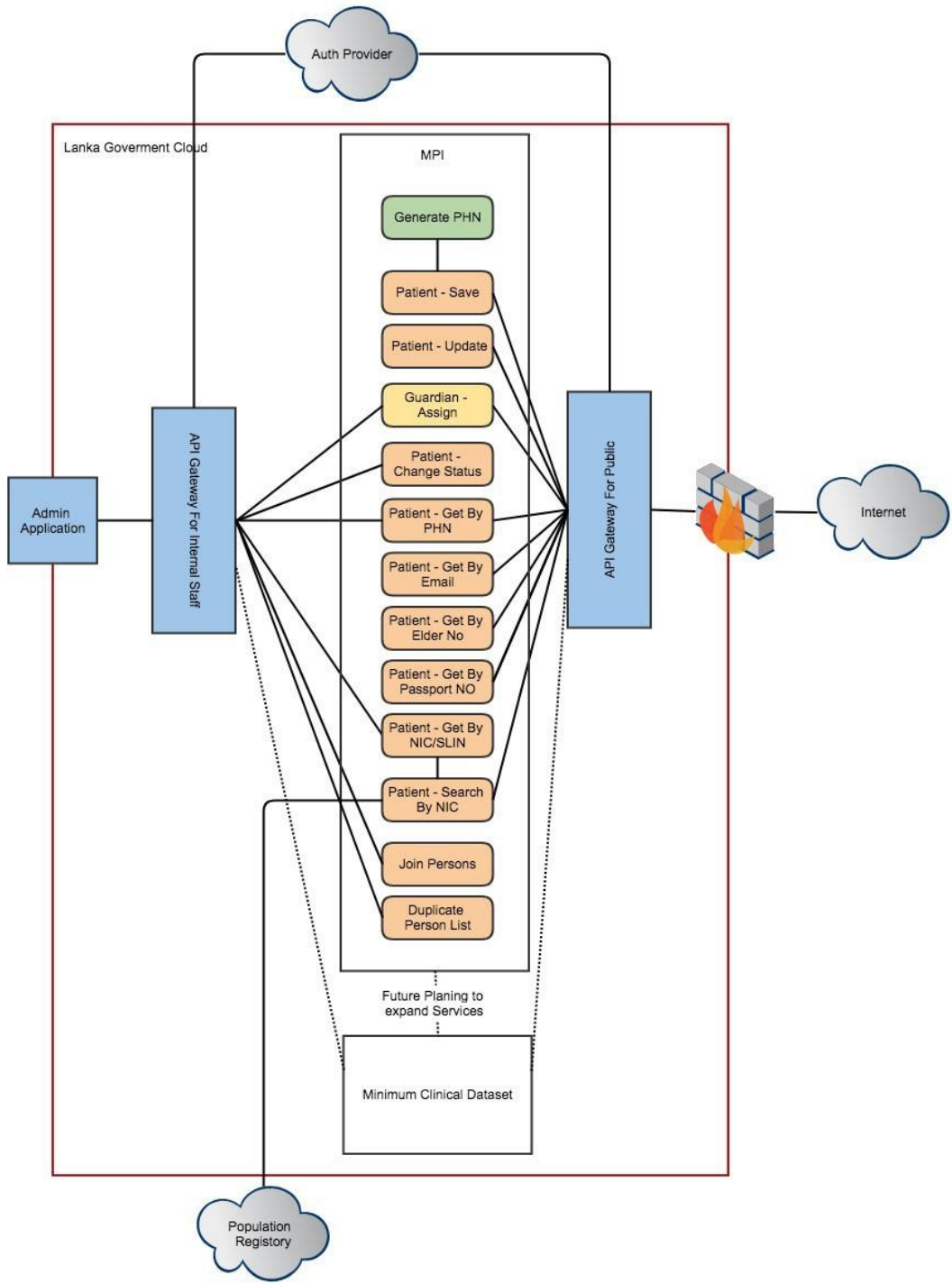

Figure 3. 3 MPI Architecture Final Version

# 4. DISCUSSION

Considering the interoperability is an utmost important requirement in Sri Lanka for the future digital health as the main factor for the future digital health carrier, discussing simple example school van accident with the failure of the interoperability.

In technical interoperability failure, doctor of the primary care unit of a base Hospital receives an electronic message from a doctor of peripheral unit, a critically injured student has no allergies to drugs. However, an electronic message was changed during transmission, the doctor gives antibiotic and patient get an allergic reaction to the drug.

When semantic interoperability failure: Doctors of the PCU of a Base Hospital sends an electronic message to asking GP of student home town where the patient clinical records keep. whether there were any warnings about a visiting student's health. The doctor from the home town sends the message code "N/A" that asserts that the information regarding allergies was never gathered. The doctors of the PCU of a Base Hospital interprets the response as "Negative for Allergies" and gives an antibiotic drug, the patient get an allergic reaction to the drug.

When process interoperability: the van load of injured schoolchildren taken to a remotely located peripheral unit. Peripheral hospital health information system does not show that the hospital's antibiotic supply is over months back. Student failure to receive antibiotic drugs.

Information communication is a key component in any system. In the health area, health information transfer among health professionals, hospitals. Effective communication required that health information handlers to share a common reference framework that allows health care information handlers interaction. So, this need a common standard. The common standard that provides the provide a common framework will promote uniformity in the health information identification in the health care system. These standards should be relatively straightforward and inexpensive in the long term(20).

However, these standard solutions are very much specific use. It is impossible adapt these standards to new arising problems, because they have given specific answer to specific questions that arises in the use of the systems. These are   more complex to implement, but standard can be adapted or scaled to different kind of scenarios, providing the essential framework(21).

According to the classification of the Angelo Rossi the computer based clinical data handling follows under the branch of semantic interoperability and to make the continuation of care the semantic interoperability should be maintained. Health data printed on the paper and displayed as the layout –based presentation. In that format health data will be saved in a repository but that will not be useful unless some person searches it.The clinical correlation is needed so the next step is present health data in a way that correlate with the clinical finding like a health care aware presentation(22). There are multiple ways that the data could be search for the repositories. With the interoperability and interconnection among the systems will create multiple ways to search the health data through the same institution or different institutions. Using the standard and guild lines will enable the data flow more smoothly while the technical and process interoperability will follow the path. At the end update of the existing data and the newly generated data will be in the system.

## 4.1 Development of Prototype

With the introduction of the HHIMS version 2 identification number for the patient were given at the institution with the established HHIMS. But it was not an unique number. 2015 onwards ICTA planning for unique identification method and introduced the PHN with collaboration of the health ministry. During discussion Master Patient Index topic came to discussion and both health care domain and the technical domain agreed to develop a MPI for Sri Lanka.

ICTA produced me the software developer team. Considering the work, they have done in other filled their technical background were solid. In lay terms, they know their staff well. A greater team is formed with the contribution of the two domains at the same time. By handling the both health domain and technical domains given opportunity to adapted to build a successful MPI.

After understanding the basic requirement of the MPI formulate a research question and the research project proposal which was accepted by the board of studies at the PGIM. The ethical clearance taken from the ethical review committee of the PGIM. After discussion with supervisor the methodology form. ADR method is suggested for the development of the MPI. Phase 1 focused on the requirement gathering from the HHIMS/HIMS users in two districts and from the and semi structured interviews from the software developers. After ending all discussions in the phase 1 following conclusions were made.

As Searching criteria for patient identity.

1. NIC
2. PHN
3. Driving license Number
4. Passport Number
5. Elderly number
6. Email
7. Name

Then children, patient with unsound mind admitted to the hospital will be assign a guardian. But new born who are discharge from the hospital after implementation of the MPI will be provided a PHN for the further used. So, till the backlog of the children who had born before the implementation will be assign to the guardian.

Elderly number consider as a good criterion od searching because when come to loss of NIC in elderly people. Email consider if available with the patient. When come to name as the criteria of searching FGD participants as the least priority towards the unique identification method.

Status update consider as inactive for the death patient. But administrator of the hospitals suggested to keep the death record for future use in case of judicial purposes for 5 years. Suggestion were made the duplicated records should be analyzed by the both personals by the health domain and technical domain to merged to single PHN number. Algorithm suggested for the merging of the duplicate to single PHN.The discussion about the algorithm will go beyond the scope of my research area and it is based on the mathematical formulas.

According to ADR method during the phase 2 focused on the designing of the MPI architecture. During the discussion done with the developer team the come with alpha version for the MPI design architecture. But the security is considered as the utmost important thing of the MPI so the from the alpha version to change to the beta version of the Architecture. When considering all other background possibilities, the last final version build.

Final phase of the development of the MPI is mainly the technical. In the development of the API for the MPI done after doing lot of research on literature review of MPI done in the history. As the message method of the XML used in the form of HL7 version 2.

## 5. CONCLUSION AND RECOMMENDATION

### 5.1 Conclusion

MPI is the 1st step of the interoperability in HHIMS/HIMS in Sri Lanka and continuation of care pathway by multiple care providers. Enable exchange of minimum essential data set among different HHIMS/HIMS systems for Sri Lanka. With the introduction of PHN will integrated the LIS, PACS, Citizen mobile application to the public access. After development of MPI prototype was implemented for national use. PHN link with all the available system in the health domain. Expand can done through private sector by providing the API publications.

### 5.2 Limitations

This MPI version is the early version main need to refine the things with the later versions. As this is a new thing to HHIMS user need more time to understand the possible questions asked by the user and problem with use of the MPI in hospitals. Private sector and preventive sector were not included and only the curative sector is used.

### 5.3 Future work

Clinical repository or MCDS will be integrated to the PHN in the version 2 of the MPI. Retrieval of Past Medical History, Allergy History. Family History from multiple EMRs to the EMR at current care provider. Need to test in preventive and private sectors and need longer period of observations for a profound understand of possible issues. Development of the citizens mobile application will help the patient s to track their clinical record online. This will be helpful in the future for continuation of care because of the tracking of the caregiver information and the practices will prevent malpractices by doctors.

### 5.4 Contributions

With this research, I contributed to the development of master patient index through making the requirement document through FGD. Assisted to develop the MPI architecture. Literature review identify a tool for evaluation of the architecture. Complete design of the HL7 message templates and PHN incorporation.

# REFERENCES

[1]   M.S. Fetter, Interoperability—Making Information Systems Work Together, Issues in Mental Health Nursing, (2009). [cited 2017 Mar 22]; Available from:http://www.tandfonline.com/loi/imhn20 30:7, 470-472, DOI: 10.2009

[2]   HL7. Health Level Seven® INTERNATIONAL [Internet]. [cited 2017 Mar 30] Available from: http://www.hl7.org/

[3]   B.H Just, D. Marc, M. Munns, and R. Sandefer, Why Patient Matching Is a Challenge: Research on Master Patient Index (MPI) Data Discrepancies in Key Identifying Fields. [cited 2017 Mar 22]; Available from: https://www.ncbi.nlm.nih.gov/pmc/articles

[4]   Enterprise master patient index - Wikipedia [Internet]. [cited 2017 Mar 22]. Available from: https://en.wikipedia.org/wiki/Enterprise_master_patient_index

[5]   L.C Burton, G.F Anderson, and I. W Kues. Using Electronic Health Records to Help Coordinate Care. [cited 2017 Mar 22]; Available from: https://www.ncbi.nlm.nih.gov/pmc/articles/PMC2690228/

[6]   G. Morris , G.Farnum, Patient Identification And Matching Final Report.[cited 2017 Mar 29] Available from: http://journal.ahima.org/wp-content/uploads/ONC-Patient-Identification-MatchingFinal-Report-February-2014.pdf

[7]   Sri Lanka - Provincial Councils [Internet]. [Accessed 2017 May 5]. Available from: http://www.priu.gov.lk /ProvCouncils/ProvicialCouncils.html

[8]   Medical Statistics Unit. Annual Health Bulletin - 2012. Ministry of Health; 2012.

[9]   P.G.N Arzt, G.D.T.Flewelling, T.J. D. Kamens, M.Rozen, and S.S Jean. Health Level Seven EHR Interoperability Work Group February 7, 2007. [cited 2017 April 9] Available from: http://www.hl7.org/special/committees/ehr/index.cfm

[10]  Maung K. Sein. ACTION DESIGN RESEARCH. Sein, Maung K.; Henfridsson, Ola; Purao, Sandeep; Rossi, Matti; and Lindgren, Rikard. 2011. "Action Design Research," MIS Quarterly, (35: 1) pp.37-56

[11]  T.Snyder, A.P Honey.An Architecture-Centric and Ontology-Based Approach to Cross-Domain Interoperability of Health Information Systems for Diabetes Care [cited 2017 Mar 24]; Available from: http://www.researchandinnovationbook.com/PROCEEDINGS/CITA2015/Archives/papers/SD3.pd

[12]  Snomed International. SNOMED-CT [Internet]. Available from: www.snowmed.org

[13]  Building a better delivery system: A New Engineering/Health Care Partnership, The National Press,2005. [Internet]. Available from: http://newton.nap.edu/book/030909643X/html

[14]  H. B. Newcombe, J. M. Kennedy, S, J. Axford, A. P. James. Automatic linkage of vital records "Computers can be used to extract "follow-up" statistics of families from files of routine records." [cited 2017 Mar 22]; Available from: https://www.cs.umd.edu/class/spring2012/cmsc828L/Papers/Newcombe59.pdf

[15] Larry & C. Fleming. Design and implementation of a data base for medical records. [cited 2017 Mar 22]; Available from: https://www.ncbi.nlm.nih.gov/pmc/articles/PMC2203794/

[16] Tan LT, National Computer Board, Republic of Singapore. National Patient Master Index in Singapore. [cited 2017 Mar 22]; Available from:Int J Biomed Comput 1996 Jan;40(3):241-2.

[17] D.W. Forslund. Federation of the personal identification services between enterprises. [cited 2017 May 02]; Available from: https://www.ncbi.nlm.nih.gov/pubmed/?term=Federation+of+the+personal+identification+services+b etween+enterprises.

[18] Joffe E, Bearden CF, Byrne MJ. Duplicate patient records—implication for missed laboratory results. [cited 2017 May 02] . Proc AMIA Annu Symp 20121269–75.

[19] S.Jain. A Prototype Automated Resolution Service for Public-Health Master Person Index. [cited 2017 May 12] Available from:http://digitalcommons.usu.edu/gradreports/239

[20] C.Toth, E.Durham, M.Kantarcioglu, Y.Xue and B. Malin . SOEMPIA Secure Open Enterprise Master Patient Index Software Toolkit for Private Record Linkage. [cited 2017 May 22]; Available from https://www.ncbi.nlm.nih.gov/pubmed/25954421

[21] M. Lee, E.Heo, H.Lim, J.U Lee, , S. Weon, H Chae, H.Hwang, S. Yoo, Developing a Common Health Information Exchange Platform to Implement a Nationwide Health Information Network in South Korea. January21. [cited 2017 May 22]; Available from:Healthc Inform Res. 2015 Jan;21(1):21-9. doi: 10.4258/hir.2015.21.1.21. Epub 2015 Jan 31.

[22] A.Pakalapati. A Flexible Consent Management System for Master Person Indices. [cited 2017 May 22]; Available from: http://digitalcommons.usu.edu/cgi/viewcontent.cgi?article=1193&context=gradreports